****** 7.10.2016 ********** 10.10.2016 ******11.10.2016***
*****13.10.2016****22.10.2016****16.11.2016***17.11.2016

\documentclass[12pt]{article}
\usepackage{blois,graphicx}

\newcommand{\href}[2]{ \, #2}

\begin{document}

\title{Average transverse momenta of baryon production at p-p collider experiments and their crucial implications for the high energy hadroproduction physics}

\author{Olga Piskounova}

\address{P.N.Lebedev Physics Institute of Russian Academy of Science, Leninski prosp. 53, 119991 Moscow, Russia}

\maketitle\abstract{ 
The phenomenological approach in the framework of Quark-Gluon String Model (QGSM) has been applied to the description of transverse momentum spectra for the various sorts of baryons in the multi particle production at modern colliders. 
The analysis of data on hyperon spectra, $dN/dp_t$, in the region 0.1 GeV/c $< p_t <$ 3 GeV/c from many experiments (ISR, STAR, UA1, UA5 and CDF) observes an important difference in the dynamics of multiparticle production in proton-proton vs. antiproton-proton collisions. From the point of view of the QGSM, the significant contribution to particle production spectra in antiproton-proton reactions goes from the antidiquark-diquark string fragmentation of pomeron diagram. This effect takes place at the intermediate energies untill one pomeron exchange is dominating. The complete study of the energy dependence of baryon average transverse momenta for the contemporary collider experiments at the energies from $\sqrt{s}$=0.2 to 7 TeV shows the slight growing of $<p_t>$ with energy. No dramatic changes were seen in the hadroproduction characteristics at the energies in the range from Tevatron to LHC, which would be responsible for a "knee" in the cosmic ray proton spectra. Thus, the phenomenon has an astrophysical origin. The average transverse momentum analysis through the different mass of hadrons reveals some regularity in the mass gaps between heavy-quark-hadron generations. This observation gives the possibility to suggest more hadrons with the following masses  13.7, 37.3, 101.5, 276, 750 ... GeV that are produced by geometrical progression with the mass factor of order $\delta{lnM}$=1. This hadrons either have new quantum numbers or are the metastable multi quark states. The QGSM ability to construct the spectra of various baryons in the entire range -1$< x_F <$1 gives advantages for the explanation of many types of asymmetries and ratios caused by hadroproduction: the nonzero baryon/antibaryon asymmetry in the central region of proton-proton collisions, the baryon-to-meson abnormal ratios in nucleus-nucleus reactions, the growing ratios of secondary particles in cosmic ray spectra as well as the recently observed negative asymmetry for charmed baryon/antibaryon spectra  at the rapidity point Y=2 in LHCb experiment. The positive asymmetry between spectra of baryons and antibaryons at the central rapidity can be only explained by the contribution from the String Junction transfer that brings the positive baryon charge from beam protons up to the region of central rapidity. Specific form of baryon production spectra causes also the diffusion of extra baryons from diquark fragmentation area of spectra into the low $p_t$ region at the nucleus-nucleus reactions. The leading character of baryon spectra in the region of target fragmentation becomes apparent in the antiparticle-to-particle ratios, which are growing with energy in the laboratory system of cosmic matter interactions. The negative baryon/antibaryon asymmetry at nonzero rapidity is the result of interplay of central baryon production and leading production in the kinematical region of diquark fragmentation. This study is a good example how the routine analysis of collider data gives many impressive implications for the interpretation of broad variaty of hadron physics phenomena. 
}

\section{Introduction}

The transverse momentum distributions of hadrons are the very primary data that have been obtained in the hadroproduction studies at the modern colliders. The baryon transverse momentum spectra are sensitive to the asymmetrical reactions, like antiproton-proton collisions. The spectra of baryons at LHC can explain the features of cosmic ray particle spectra at very high energies: the average transverse momentum dependence on energy should show the special points, where the characteristics of hadroproduction happend to be changed due to some important reasons. The mass dependence of average transverse momenta up to heavy flavor hadrons can demonstrate some tendencies at the comparison of light and heavy quark hadron generations. The leading/nonleading asymmetry in the spectra of baryon production at the proton fragmentation regions causes the visible growing antiparticle-to-particle ratios of secondary's in the laboratory system at the measurements in cosmic space. This is a short list of statements that have motivated me to study the baryon transverse momentum spectra on the wide range of proton collider energies.

\begin{figure}[htpb]
  \centering
  \includegraphics[width=8.0cm, angle=0]{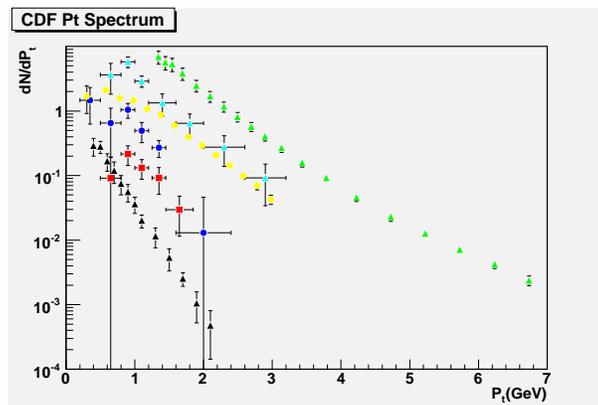}
  \caption{ Transverse momentum distributions from different experiments and of various energies. The data are from STAR(200GeV)- black triangles; UA5 energies: $\sqrt(s)$ = 200 GeV(red squares), 546 GeV(blue circles) and 900 GeV(aqua triangles); UA1(630GeV)- yellow circles and CDF(1.8 TeV)- green triangles.}
  \label{spectracompiled}
\end{figure}

The compilation in figure~\ref{spectracompiled} illustrates the changes in baryon transverse distributions on the energy distance from ISR to LHC experiments. The low $p_t$ area 0.3 GeV/c $< p_t <$ 8 GeV/c is responsible for the value of average $p_t$. We see that this value has to grow with energy.

Interpretation of these distributions in the up-to-date phenomenological models can shed light on the physics of hadroproduction processes at high energies. The Quark-Gluon String Model [QGSM] approach \cite{qgsm} is applied here to the description of $p_t$ spectra for all available flavors of baryons. 
The Model described long ago the big volume of data from previous generation of colliders  up to the energies $\sqrt{s}$= 53 GeV in the area of low $p_t$'s \cite {veselov}. 
Recently $\Lambda^0$ production have been studied \cite{lambda} in updated version of QGSM.  
 
The analysis of data on hyperon transverse momentum distributions, $dN/dp_t$, that were gathered from various experiments (ISR, STAR, UA1, UA5 and CDF) reveals an important difference in the dynamics of multiparticle production in proton-proton vs. antiproton-proton collisions in the region of transverse momenta 0.3 GeV/c$ < p_t < $3 GeV/c. Hyperons produced with proton beam display a sharp exponential slope at low $p_t$, while those produced with antiproton beam do not. Since LHC experiments have proton projectiles, the spectra of multiparticle production at LHC should seem softer in comparison to the expectations, because the MC predictions were based on Tevatron (antiproton-proton) data. From the point of view of the Quark-Gluon String Model, the most important contribution to particle production spectra in antiproton-proton reactions is due to antidiquark-diquark chain fragmentation.

The suggested range of energy seems also interesting for the detail study of baryon spectra, because the cosmic ray proton spectrum has the "knee" at the very energy gap between Tevatron and LHC experiments \cite{knee}. The change in the slope of spectrum of protons, produced in the space, may have an astrophysical origin. Otherwise, it would mean the substantial change in the dynamics of hadron production. We have to learn the behavior of baryon production spectra moving from smaller energy to higher, in order to conclude that nothing dramatic takes place in hadroproduction processes at "knee" energy or a little above.          

It is effective to consider here the latest baryon production data from LHC experiments and to calculate the average transverse momenta of various baryons as a function of c.m.s. energy of colliding protons, $\sqrt{s}$, in order to conclude whether or not a new regime appears at the energy of "knee" in cosmic proton spectra. If this dependence has some specifics near the knee energy, we have to expect the change in the slopes of cosmic spectra due to the hadronic interactions in atmosphere or in detectors.

It is interesting as well to compile the available data on $<p_t>$ at certain LHC energy for all sorts of hadrons and to present them as a function of hadron mass. Some regular features could be observed between the hadron generations.

The previous QGSM study \cite{antiprotonratio} has shown that the charge ratios of secondary antiparticle-to-particle spectra in cosmic rays grow due to the leading production asymmetry of baryon spectra towards the antibaryon spectra in the kinematical region of target fragmentation, where targets are mostly built of the positive matter. The procedure of spectra recalculation from center-of-mass system at collider p-p collisions into the laboratory system at cosmic ray interactions was developed in \cite{gamma} and is presented in the Appendix.

\section{Preliminary comparison of hyperon transverse momentum spectra from collider experiments}

The recent data on $\Lambda^0$ hyperon distributions are obtained in the following LHC groups: ALICE \cite{alice} at 900 GeV, ATLAS \cite{atlas} and CMS \cite{cms} at 900 GeV and 7 TeV. 
We are going to compare the results of LHC groups with the data of lower energy colliders, UA5 \cite{isr}($\sqrt{s}= 53 GeV$)and STAR \cite{star}($\sqrt{s}= 200 GeV$).
  
\begin{figure}[htpb]
  \centering
  \includegraphics[width=8.0cm, angle=0]{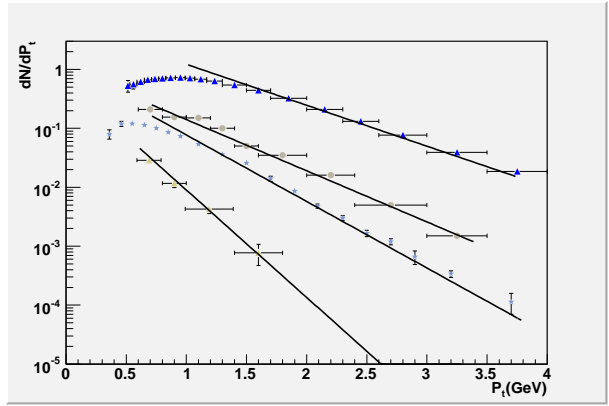}
  \caption{ Transverse momentum distributions from ISR (53 GeV)-yellow triangles, STAR (200 GeV)- blue stars, ALICE (900GeV)- grey circles and ATLAS (7 TeV - blue triangles, as fitted with the exponent.}
  \label{isr_star_alice_atlas}
\end{figure}

In figure~\ref{isr_star_alice_atlas} the flattering in the slope of $dN/dp_t$ can be seen, if we fit the data with a simple exponential function: exp(-B*$p_t$). We can conclude that transverse momentum spectra are more and more hard with the energy growth that provides the change in the slopes, beginning from B=4,2 for ISR data, B=2,6 for STAR and to B=2,0 at 900 GeV in ALICE. The slope is more flat if we take the spectra at $\sqrt{s}$ = 7 TeV , B = 1,5. This is the reason to discuss the growing of average baryon transverse momenta with energy.
 
\begin{figure}[htpb]
  \centering
  \includegraphics[width=8.0cm, angle=0]{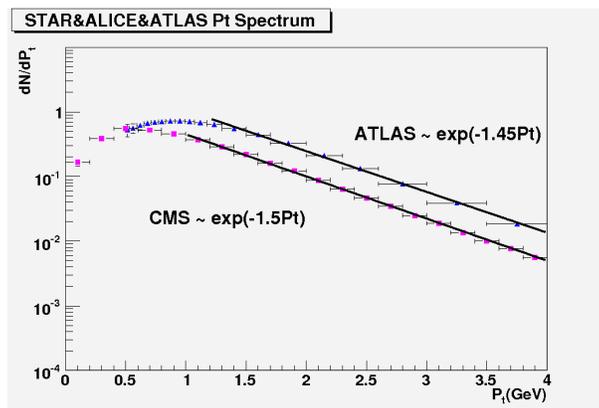}
  \caption{ Transverse momentum distributions at $\sqrt{s}$= 7 TeV: ATLAS - blue triangles and CMS - pink squares}
  \label{atlas_cms}
\end{figure}

Both LHC experiments at 7 TeV, ATLAS and CMS have presented the hyperon spectra, see figure~\ref{atlas_cms}, with the similar slopes that is expectable. The different forms of the distributions at low $p_t$ values may be caused by specifics in efficiency of the detecting procedure. It should be mentioned here that ATLAS measurement procedure has systematically low efficiency at $p_t< 1$ GeV in the comparison to the results of other LHC groups.

\section{Baryon transverse momentum distributions in QGSM}

As we have studied in early QGSM paper \cite{veselov}, the transverse momentum spectra of produced hadrons in proton-proton collisions can be perfectly described with more complicate $p_t$-dependence:

\begin{equation}
E \frac{d^{3}\sigma^H}{dx_F d^{2}p_{t}}= \frac{d\sigma^{H}}{dx_F}*A_0*\exp[-B_0*(m_t-m_0)],\nonumber 
\end{equation}
where $m_0$ is the mass of produced hadron, $m_t$ = $\sqrt{p_t^2+m_0^2}$ and $B_0$ used to bring the dependence on $x_F$, but in central region of rapidity this slope is constant. The slopes for the spectra of many types of hadrons ($\pi$, K, p, $\Lambda^0$) were estimated for the data of proton-proton collisions at the energies available those times. The value of the slopes of baryon spectra was universal and equal approximately $B_0$ = 6,0.  

\begin{figure}[htpb]
  \centering
  \includegraphics[width=8.0cm, angle=0]{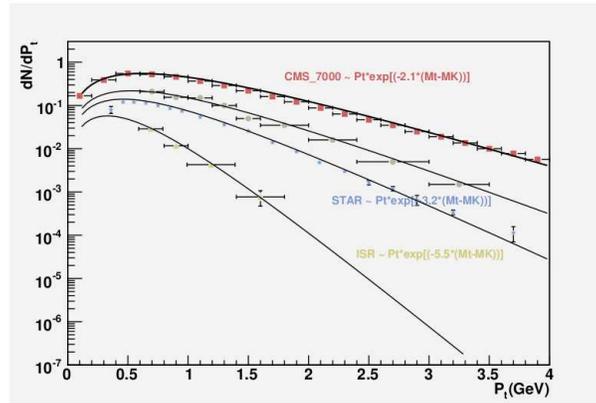}
  \caption{ The description of ISR, STAR, ALICE(900GeV) and CMS data on hyperons with the QGSM function.}
  \label{hyperonspectra}
\end{figure}

Now we have to conclude that the slopes of spectra, $B_0$, at the modern collider experiments depend on energy. More, as it is seen from the spectra at LHC and RHIC, the value of $m_0$ is not the mass of proton or hyperon.
The better description of hyperon spectra, see figure~\ref{hyperonspectra}, can be achieved with $m_0$ = 0,5 GeV that is actually the mass of kaon. This effect is suggestively explained as the minimal transverse momentum of hyperon at the fragmentation of diquark-quark chain (see the QGSM pomeron diagram for p-p collisions). It is should be equal to the kaon mass, when the minimal diquark-quark chain fragmentation gives only two hadrons: $\Lambda^0$+K.

\section{The difference between transverse momentum distributions in proton-proton and antiproton-proton collisions}

\subsection{UA5 and STAR data}

\begin{figure}[htpb]
  \centering
  \includegraphics[width=8.0cm, angle=0]{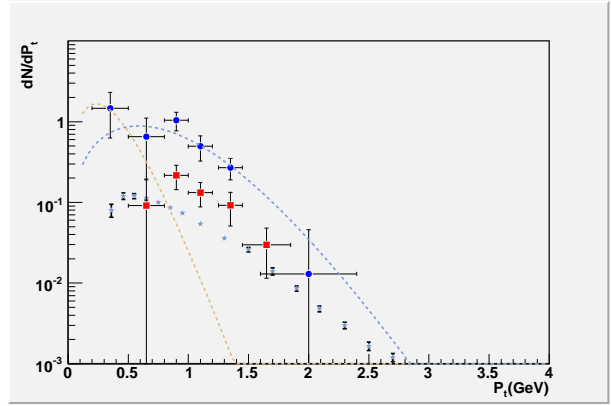}
  \caption{The forms of spectra at antiproton-proton reaction (UA5)-red squares vs. poton-proton one (STAR)-blue stars at the same energy ($\sqrt{s}$=200 GeV). We remember that the absolute values of spectra are chosen arbitrary. The UA5 data of higher energy ($\sqrt{s}$=546 GeV) are shown with blue squares. The fit with two dashed lines demonstrates two different components at the asymmetric one-pomeron cut.}
  \label{antiprotonVSproton}
\end{figure}

The data of many high energy experiments on $p\bar{p}$ collisions \cite{ua5,cdf} as well as on $pp$ collisions of lower energies \cite{star,isr} have been considered in this research in order to understand the influence of quark composition of beam particles on the shape of transverse momentum spectra of $\Lambda^0$ hyperon production, see figure~\ref{antiprotonVSproton}. Unfortunately, the difference in spectra was not studied enough at ISR energies, where proton and antiproton projectiles were both available.

Obviously, the form of the spectrum at low $p_t$ has strong impact on the value of cross section, when the experimental distributions are integrated beginning from rather low momenta, $p_t >$ 0.5 GeV/c.
The resulting cross section from antiproton-proton reaction is to be smaller than the cross section, obtained in proton-proton collision of same energy, as we will learn from the next subsection, where we explain the difference between the forms of baryon spectra in antiproton-proton experiments (UA1, UA5, Tevatron) and in proton-proton collisions (ISR, STAR and LHC experiments of various energies) with the help of QGSM one-pomeron cut diagrams. 

\subsection{The difference as it is shown in diagrams of Quark-Gluon String Model}

The difference in $p_t$-spectra of $\Lambda^0$'s produced in high energy $p-p$ and $p-\bar{p}$ collisions cannot be explained in the perturbative QCD models, because both interactions should give the mutiparticle production in central rapidity region due to Pomeron (or multiPomeron) exchange.
The total cross section and the spectra in $pp$  and $p\bar{p}$ collisions are to be equal because they depend only on the parameters of the Pomeron exchange between two interacting hadrons and should not be sensitive to the quark contents of colliding particles. 

\begin{figure}[htpb]
  \centering
  \includegraphics[width=8.0cm, angle=0]{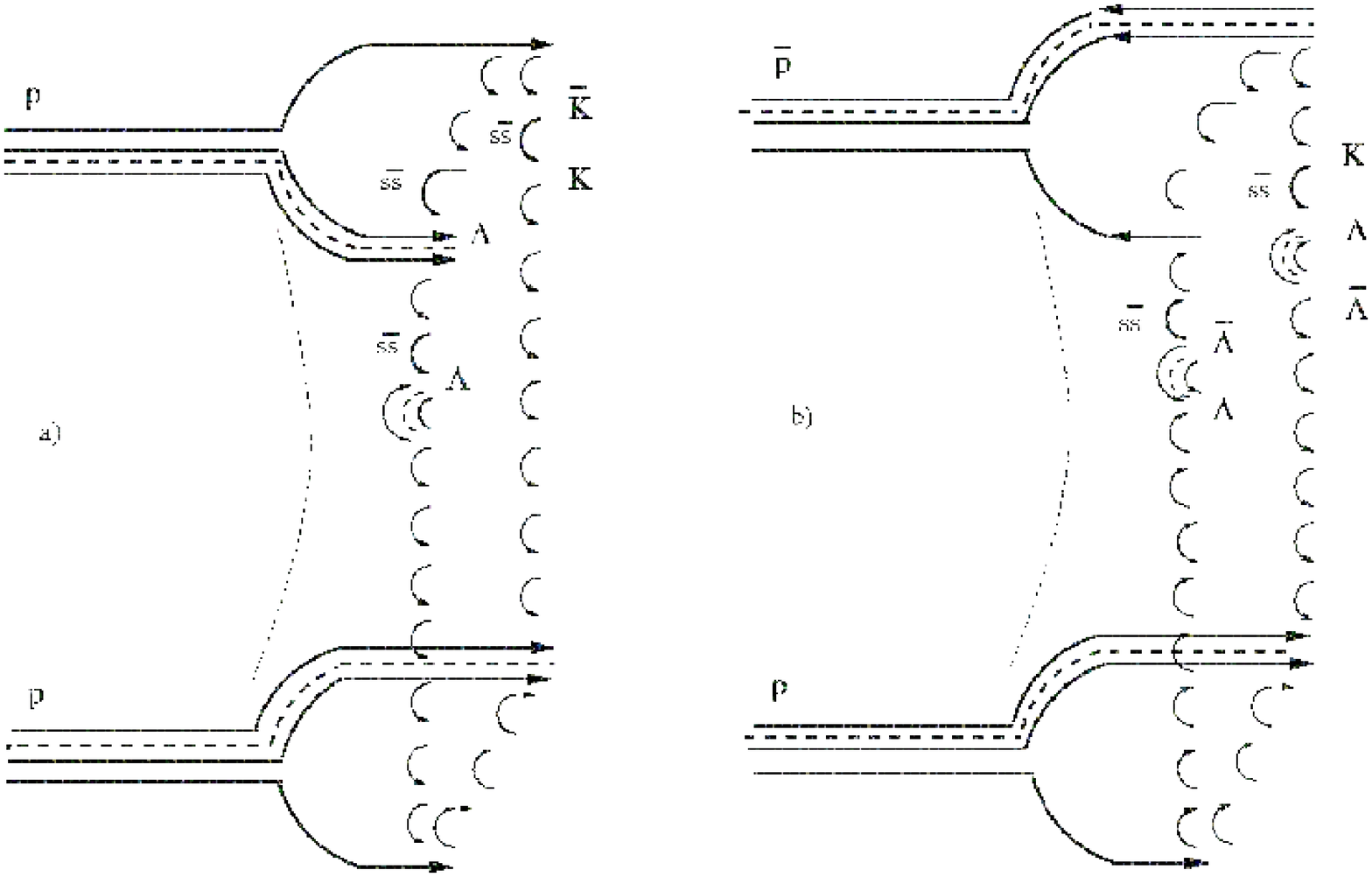}
  \caption{QGSM multiparticle production diagrams for a) $pp$ and b) $p\bar{p}$ reactions.}
\label{pomerondiagrams}	
\end{figure}

Let us look at the pomeron diagrams of proton-proton and antiproton-proton collisions that are shown in the figure~\ref{pomerondiagrams}.
From the point of view of Quark-Gluon String Model \cite{qgsm,hyperon}, which is based on the Regge theory and on the phenomenology of pomeron exchange, the spectra of produced particles are the results of the cut of one-pomeron diagrams. As it is seen from the comparison of diagrams, the most important contribution to hadron production spectra in antiproton-proton reaction is brought by the fragmentation of antidiquark-diquark chain of pomeron cylinder, because this side of diagram certainly takes the most part of energy of colliding $p\bar{p}$. Otherwise, the proton-proton collision diagram is symmetric and built from two similar quark-diquark chains. The difference exists only because of low sensitivity in the region of small transverse momenta, where the hadron spectra at antiproton-proton reaction have been underestimated. This effect is seen in the production cross sections that are systematically lower for Tevatron data in the comparison to LHC results. So, the difference in the spectra is a local effect that has no influence on the integral value of hadroproduction cross sections, which have similar pomeron (and multypomeron) asymptotical dependence on the energy.

The important fact is that the latest experiments of highest energies before LHC were carried out with antiproton beams. It seems a mistake to suggest that $pp$ and $p\bar{p}$ at high energy are to give the similar hadron transverse momentum distributions. The spectrum of hyperons that was produced with proton beam at STAR \cite{star} has the sharp exponential slope at low $p_t$, while the spectra with antiproton beam are of more complicate configuration. 

In the same time, the low $p_t$ sharp exponential contribution seams existing in hyperon spectra in $p\bar{p}$ reaction if we pay attention to the spectrum at $\sqrt{s}$ = 546 GeV in UA5 collaboration, see figure~\ref{antiprotonVSproton} \cite{ua5}. Due to the data at $\sqrt{s}$ = 546 GeV we can conclude that the sharp exponential slope is still there at very low $p_t <$ 0.5 GeV/c. It gives the hope that this exponential addition exists in other antiproton spectra as well, but it doesn't seen because of the absence of measurements at very low $p_t$'s.

In common, the difference in $p_t$-pectra at $pp$ and $p\bar{p}$ collisions has not to give any impact on the total cross sections. Our study is practically concentrated on the form of antidiquark-diquark string fragmentation in antiproton-proton collisions that results in a visible maximum in the hadron spectra at the $p_t$ region: 0.8 GeV/c $< p_t < $2. GeV/c. But if we neglect with the data points at $p_t < $0.5 GeV/c, the total cross sections for $p-\bar{p}$ colliders are underestimated.

\begin{figure}[htpb]
  \centering
  \includegraphics[width=5.0cm, angle=0]{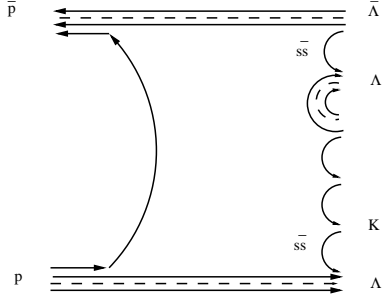}
  \caption{The low energy diagram with quark-antiquark annihilation and antidiquark-diquark chain fragmentation.}
\label{lowenergy}	
\end{figure}

The nature of the excess is in the asymmetric share of energy between the sides of pomeron cylinder and have to be studied in future proton-antiproton experiments of low energy \cite{tapas}. At low energy the annihilation between quark-antiquark takes place, as it seen from the figure~\ref{lowenergy}. The resulting spectra consists of the contribution from diquark-antidiquark chain that allows us to learn the pure form of distribution at diquark-antidiquark fragmentation.

Since LHC experiments have proton projectiles, the spectrum of hyperon production seems "softer" in the comparison to one that is predicted by Monte Carlo generators and was preliminary tuned to the Tevatron data.

\begin{figure}[htpb]
  \centering
  \includegraphics[width=8.0cm, angle=0]{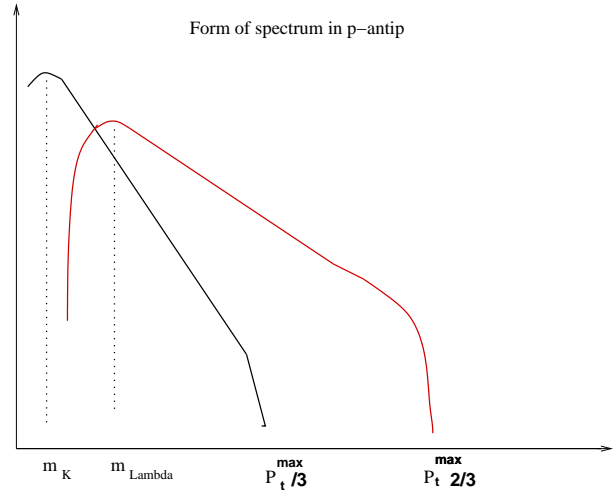}
  \caption{ The form of hyperon spectrum with two different contributions from quark-antiquark and diquark-antidiquark chains.}
\label{formofspectra}	
\end{figure}

The energy splitting in the case of symmetric proton-proton pomeron diagram should lead, on my opinion, to the exponential $p_t$ distribution that goes from low $p_t$'s of the order of kaon mass. In the same time the asymmetric share of energy in $p\bar{p}$ makes the visible maximum in the spectra at $p_t$ near the hyperon mass due to contribution of diquark-antidiquark as it is demonstrated in figure~\ref{formofspectra}.

\section{Average Baryon Transverse Momenta vs. Energy and "Knee" in Cosmic Ray Spectra}

Due to the difference between the baryon spectra in $\bar{p}-p$ reaction and in $p-p$ one, which was discussed above, we have to confine ourselves to the further consideration of only proton-proton collision experiments : STAR, ALICE, ATLAS and CMS in the wide range of $\sqrt{s}$ energies: beginning from 200 GeV to  the up-to-date LHC energy 7 TeV.

\begin{figure}[htpb]
  \centering
  \includegraphics[width=8.0cm, angle=0]{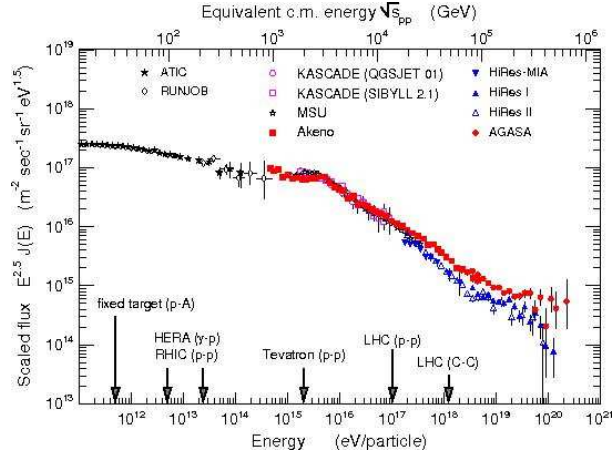}
  \caption{The cosmic proton spectrum with the "knee" between Tevatron and LHC energies}
  \label{knee}
\end{figure}

The suggested range of energy has motivated me to study the behaviors of baryon spectra, because the cosmic ray proton spectrum shows the "knee", see figure~\ref{knee}, at the very energy in between Tevatron and LHC colliders \cite {knee}. The change in the energy spectrum slope of protons, produced in space, may be of astrophysical origin, otherwise it means a substantial change in dynamics of particle production. We have to learn the behaviors of baryon production spectra from one energy point to another in order to conclude that no dramatic changes have place in hadroproduction processes at "knee" energy and above. 

\begin{figure}[htpb]
  \centering
  \includegraphics[width=8.0cm, angle=0]{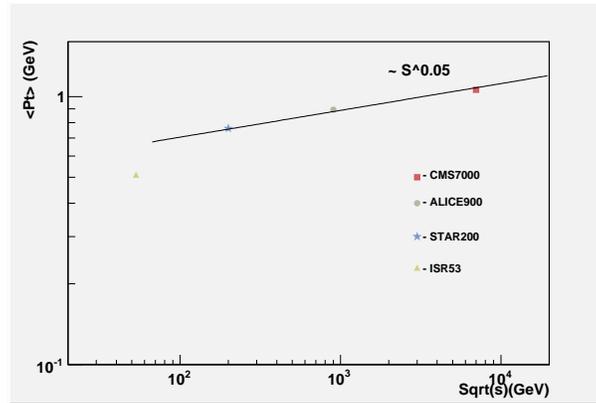}
  \caption{ The resulting values of average hyperon $p_t$ that are growing with the energy and the asymptotic dependence $ ~ s^{0.05}$.}
\label{averageVSenergy}	
\end{figure}

The average baryon transverse momenta dependence has to indicate a specifics at some energy point in the considered range. As it was learned from the previous QGSM studies, the typical average transverse momentum is almost constant or is slightly growing with energy due to the contributions of multi pomeron exchanges that simply gives larger fluctuations in the transverse momenta.
In figure~\ref{averageVSenergy} the resulting dependence of average $p_t$ on the energy is shown.

The $<p_t>$ values are growing up to the STAR energies that is approximately $\sqrt{s}$ = 100-200 GeV, and then they go with the asymptote ~ $s^{0.05}$. This behavior cannot be considered as substantial change in hadroproduction processes. This statement is very important for cosmic ray physics, where the "knee" (the change in the slope) at $E_{lab}$ $\approx$ 3* $10^{15}$ eV in cosmic proton spectra might be a manifestation of new regime in hadronic interactions. As we have discussed above, nothing drastic happends with baryon spectra up to $\sqrt{s}$ = 7 TeV that corresponds to $E_{lab}= 2,5*10^{16}$ eV. It means that the "knee" is caused by some astrophysical reason. On the other hand, the "knee" may indicate, as an example, the maximal energy of protons that are being produced in other Galaxies. But the idea of production of very high energy protons in space assumes the further detailed investigations of the baryon production dynamics in the framework of our model. As for other models, the energy dependence of $<p_t>$ was not yet studied and incerted into Monte Carlo generators. Such a way, this reseach is very advantageous for the improvement of LUND, Pythia and etc., which are used in the prediction and interpretation of results of LHC-groups measurements.

\section{Average Transverse Momenta vs. Mass of Hadrons}

\begin{figure}[htpb]
  \centering
  \includegraphics[width=8.0cm, angle=0]{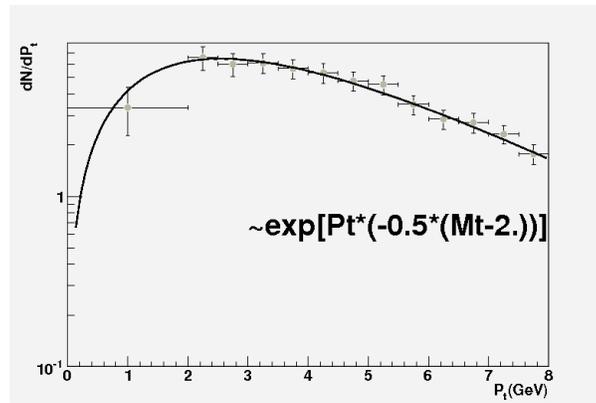}
  \caption{The QGSM fit of B-meson spectra at $\sqrt{s}$=7 TeV in LHCb experiment.}
\label{HQspectra}	
\end{figure}

The previously published analysis of transverse momentum spectra of baryons from LHC experiments (ALICE, ATLAS, CMS)\cite{bylinkin} is supplemented with the spectra of kaons, D-mesons and B-mesons from LHCb \cite{lhcbmeson}. The heavy quark meson spectra at 7 TeV were fitted with the same formula as the baryon spectra (see figure~\ref{HQspectra}).

\begin{figure}[htpb]
  \centering
  \includegraphics[width=8.0cm, angle=0]{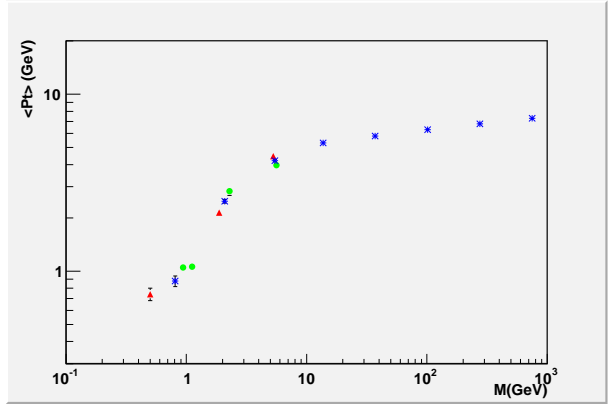}
  \caption{The mass dependence of average momenta of hadrons (mesons - green triangles, baryons - red circles). The extension for the expected heavy hadron states is shown with blue stars.}
\label{averageVSmass}	
\end{figure}

The dependence of average transverse momenta on the mass of hadron in the figure~\ref{averageVSmass} shows $<p_t>$'s that are almost-linear-growing with masses up to botom quark mass. If we imagine the point between meson and baryon masses of given quark flavor, which is "symmetric" in sense of spin (nor meson, nor baryon) - the gaps from one flavor mass to another go with the mass factor $\delta{lnM_{H}}$ = 1. It means that we have the geometric progression for the suggested hadron masses.

The easy extension of this sequence gives us the states with the following masses: 13.7, 37.3, 101.5, 276, 750 GeV and so forth. 
The similar supersymmetric hadron states were studied in \cite{rhadron}. The plot is expanded with imaginary $<Pt>$ dependence $ ~ M_{H}^{0.1}$ , because it is seen that the linear growing of average momenta is obviously stopping at bottom hadron masses. Unfortunately, the mass of top quark doesn't match this collection of possible hadron states. Suggestively, the hadrons are bringing new quantum numbers or are heavy multi quark states. But this hypothesis is not only attempt to predict new particles. It rather has to be considered as possibility to comprehend the unification of baryon and meson features at heavy masses, as well as to imagine the levels of some QCD string potential that are including, by the way, the existing hadron generations.    

\section{The Influence of Leading Baryon Production on the Charge Asymmetries and Form of the Cosmic Ray spectra}

\begin{figure}[htpb]
  \centering
  \includegraphics[width=8.0cm, angle=0]{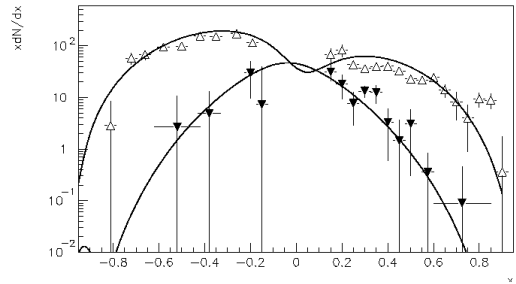}
  \caption{The form of charmed baryon spectra in $\Sigma-p$ collision for the whole $x_F$ region (the proton target fragmentation takes the left side of the distribution). The data for baryons (empty triangles) and antibarions(black triangles) are compilated from E781 experiment with the arbitrary absolute values at central point.}
\label{sigmap}	
\end{figure}

The noteworthy feature of the baryon production in proton-proton interactions is valuable asymmetry between antibaryon and baryon energy distributions in the kinematical regions of colliding beams  fragmentation that is seen in the figure~\ref{sigmap}, where the spectra of $\Lambda_c$ are described in QGSM for the entire kinematical $x_F$ range \cite{sigmap}.

Unfortunately,  the specific form of baryon spectra are often ignored in the comparisons of baryon and meson multiplicities. In the nucleus collisions the effect of "baryon anomaly" (the abnormal ratios of baryon to meson, like p/$\pi$, $\Lambda^0$/K etc. compared to those produced in proton-proton reaction) can be explained by the diffusion of additional baryons from the area of high $x_F$ of the spectra (see figure~\ref{sigmap}) into the intermediate part of transverse momentum spectra due to the secondary interactions \cite{extrabaryons}.

Large asymmetry appears as well in the spectra of protons and antiprotons in the area of high $x_F$ at the beam proton fragmentation \cite{hyperon}. Significantly, there is a dip between the growing central part of distribution (a "table") and the stable proton fragmentation region (diquark "wing"). This dip causes also the negative asymmetry between spectra of heavy baryons and antibaryons that was measured recently at LHCb experiment on some distance from the central rapidity \cite{lhcb}. Because the detail concideration of quark-diquark chain fragmentation in QGSM reveals the additional (1-$x_F$) factor in the central production of baryons as regards to the antibaryon spectra. It gives the tiny antibaryon excess for the rapidities close to the center, where the diquark fragmentation interferes with the central production. This contribution should be more visible for heavy flavor baryons. 

\begin{figure}[htpb]
  \centering
  \includegraphics[width=5.0cm, angle=0]{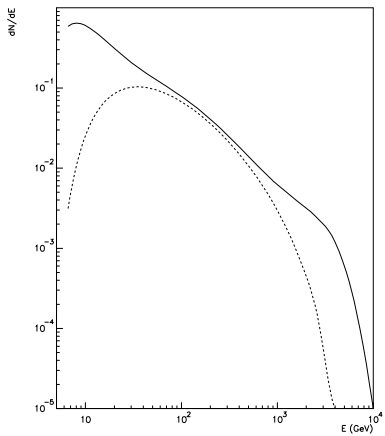}
  \caption{The spectra of $\Lambda_c$ (solid line) and anti$\Lambda_c$ (dashed line) as they were recalculated for the laboratory system of proton-proton collisions.}
\label{Lambdalab}	
\end{figure}

In the laboratory system of cosmic ray interactions, the spectra are to be converted into the energy distributions like those shown in figure~\ref{Lambdalab}, where the valuable baryon/antibaryon asymmetry takes place at the energies up to hundreds GeV. The procedure of spectrum recalculation at the transformation from the center-of-mass system into the laboratory system of coordinates was developed in paper \cite{gamma} and is represented in the Appendix.

The growing ratio between antiparticle and particle spectra that appears in the left side of spectra due to influence of positive proton target was already described in the early publication \cite{antiprotonratio}.
This suggestion makes significant the formation of cosmic ray spectra due to the particle production at the sources of very high energies.

Asymmetries in baryon/antibaryon production in the central rapidity region will be discussed in the next publication.

\section{Conclusions}

The review of results in transverse momentum distributions of hyperons that are produced in proton-proton collisions of various energies \cite{recent} has revealed a notable change in the slopes of spectra in the region of $p_t$ = 0,3 - 8 GeV/c. The spectra of baryons are becoming harder and harder with the energy growth from ISR and RHIC ($\sqrt{s}$=200 GeV) up to LHC (0,9 and 7 TeV). The detailed analysis of hyperon spectra, which is analogous to our early study in the framework of Quark-Gluon String Model, demonstrates the change of slopes from $B_0$ = 4,6 (ISR at 53 GeV) to $B_0$ = 2,1 (LHC at 7 TeV). 
The transverse momentum baryon spectra in proton-antiproton collisions (UA1, UA5, CDF) differ from the $p_t$ distributions of baryons in proton-proton collisions (ISR, STAR, LHC).
Quark-Gluon String Model explains this phenomenon as a difference in the splitting of transverse energy between two sides of pomeron multiparticle production diagram for one considered reaction vs. another. The diagram for proton-antiproton case includes the unusual string with the diquark-antidiquark ends, which certainly accumulates more energy than another quark-antiquark side of cylinder. It is natural as well to suggest that the difference in spectra will disappear with the growing of energy due to the growing multipomeron contributions into the differential cross section that are similar as for $p\bar{p}$ as for $pp$ collisions.
In order to confirm the theoretical suggestions that are described above, the measurements of hyperon transverse momentum spectra from all LHC experiments have been analyzed. The best way to speculate on the features of hadroproduction at high energies is the comparisons of average transverse momenta at various energies and various masses of produced hadrons.
First of all the energy dependence of hyperon average $p_t$ was investigated. These average values are slowly growing with energy with the power low ~$s^{0.05}$ up to highest LHC energy.

Thus it makes us conclude that processes taking place in baryon production at the up-to-date energies of LHC are not something unpredictable. This statement is very important for cosmic ray physics, where the "knee" (the change of the slope) at $E_{lab}$ $\approx$ 4* $10^{15}$ eV in cosmic proton spectra might origin in hadronic interactions. As we have discussed above, nothing dramatic happens with baryon spectra up to $\sqrt{s}$ = 7 TeV corresponding to $E_{lab}= 2,5*10^{16}$ eV. It means that the "knee" is caused by astrophysical reasons. On my mind the "knee" may indicate the maximal energy of protons that are being produced in the nearest Galaxy. But the idea of proton production in space assumes a further detailed QGSM investigation of the proton production dynamics. 
The average transverse momentum analysis through the different mass of hadrons allow to suggest some regularity in the mass gaps between heavy quark baryon-meson generations. This observation gives the possibility for more hadrons with the masses ~ 13.7, 37.3, 101.5, 276, 750... GeV that are produced by geometrical progression with the mass multiplier of order $\delta{lnM}$=1. These hadrons may possess new quantum numbers or consist of heavy multi quarks. At least this is a challenge for QCD theory.
This reseach is very advantageous for the improvement of MC generators, such as LUND, Pythia and etc., which are used for the prediction and interpretation of results of LHC-groups measurements. Taking into account the great difference between analytical and Monte Carlo approaches, I hardly imagine how our baryon spectra in the beam fragmentation region can be inserted into MC generators listed above.    
The baryon/antibaryon asymmetries that were already measured in LHC experiments are also the goal for further studies. 
The growing antiparticle/particle ratios cannot be the result of interstellar matter acceleration. Growing charge ratios of secondary particle spectra in CR have resulted from the baryon production in the CR interactions with positive matter targets.
The studies of baryon-to-meson ratios in nucleus-nucleus reaction \cite{extrabaryons} have to use the QGSM baryon spectrum calculations in order to explain the registered "baryon anomaly" in nucleus-nucleus interactions. The leading form of baryon spectra can provide the extra baryons migrated from the diquark fragmentation area of spectra into the intermediate $p_t$ region due to double interactions in nuclear environment. The negative baryon/antibaryon asymmetry at nonzero rapidity is the result of interplay of central baryon production and leading production in the diquark fragmentation. These broad variety of QGSM applications would be useful for the future common works of phenomenologists and experimentalists.

\section{Appendix}
\begin{figure}[htpb]
  \centering
  \includegraphics[width=12.0cm, angle=0]{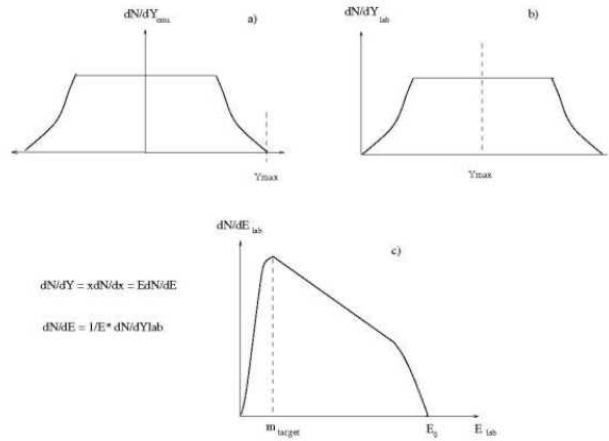}
  \caption{The step-by-step procedure of spectra recalculation from c.m.s. to laboratory system of coordinates as it was applied in the early paper.}
\label{appendix}	
\end{figure}

It was found in an earlier paper \cite{nikitin} that the typical rapidity distributions of hadrons at collider experiments can easily converted into dN/dE energy spectra in the laboratory system, where one beam particle becomes the target. This procedure is graphically illustrated in the figure~\ref{appendix}. 

This method has been developed in \cite{gamma}.
As we know the rapidity distributions, dN/dY, at high energy proton collisions looks like the "table". Rapidity spectrum in laboratory system is obtained by the shift of the "table" on the value of Ymax, so that all rapidity range is positive as in the fixed target collisions. Then our distribution should be expressed in energy variable $x$ = E/$E_{collision}$: dN/dY=xdN/dx=EdN/dE. It is clear that the energy spectrum in the laboratory system will be power like and dN/dE will be propotional to 1/E.

\section{Acknowledgments}
 
I would like to thank the Russian Foundation of Fundamental Research (grant 13-02-06091) and Dmitry Zimin's "Dynasty" Foundation. Their financial supports made possible the First Kaidalov's Phenomenology Workshop. This workshop has given the boost to the phenomenological research that is described in my paper.


\section{References}

\begin{thebibliography}{99}
\bibitem{isr} ISR Collaboration, D.Drijard {\it et.al.}, Z.Phys. C{\bf 12}, 217,1982.
\bibitem{star} STAR Collaboration, B.I. Abelev {\it et.al.}, Phys. Rev. C{\bf 75}, 064901,2007.
\bibitem{ua1} UA1 Collaboration, G. Bocquet {\it et.al.}, Z. Phys. C{\bf 366},441,1996.
\bibitem{ua5} UA5 Collaboration, R.E. Ansorge {\it et.al.}, Nucl. Phys.B{\bf 328},36,1989.
\bibitem{cdf} CDF Collaboration, D. Acosta {\it et.al.}, Phys. Rev. D{\bf 72}, 052001,2005.
\bibitem{alice}ALICE Collaboration, EPJ C{\bf 71},1594,2011, e-print - arXiv:1012.3257.
\bibitem{atlas}ATLAS Collaboration, Aad {\it et.al.},  PR D85,012001, e-print - arXiv:1111.1297.
\bibitem{cms}CMS Collaboration, Khachatryan {\it et.al.}, JHEP 05,064, e-print -arXiv:1102.4282.
\bibitem{gamma}O.I. Piskunova (Lebedev Inst.), Sov.J.Nucl.Phys. {\bf 51} (1990)846.  
\bibitem{knee}V. B. Petkov, O. D.Lalakulich, G. M.Vereshkov, Phys.Atom.Nucl.{\bf 66},523, 2003; Jorg R. Horandel, N.N.Kalmykov, A.V.Timokhin, Jou.of Phys.: Conference Series 47 (2006) 132.
\bibitem{veselov}A.I. Veselov, O.I. Piskunova, K.A. Ter-Martirosian, Phys.Lett.B{\bf 158}, 175, 1985. 
\bibitem{qgsm}Quark-Gluon String Model, A.B.Kaidalov and K.A.Ter-Martirosyan, Sov.J.Nucl.Phys.{\bf 39}, 1545, 1984, {\bf 40}, 211, 1984, A.B.Kaidalov, Phys.Lett.B{\bf 116}, 459, 1982.
\bibitem{hyperon}A.B. Kaidalov and O.I. Piskunova Z. Phys. C{\bf 30},145,1986, O.I. Piskounova Nucl.Phys.Proc.Suppl.  93, 144, 2001, e-Print: hep-ph/0010263 and Phys.Atom.Nucl.{\bf 70}, 1107, 2007, e-Print: hep-ph/0604157.
\bibitem{lambda} O.I.Piskounova, e-Print - arXiv:1209.6214.
\bibitem{tapas} Giorgio Apollinari (Fermilab), Olga Piskunova et al., FERMILAB-TAPAS-PROPOSAL-1022, Fermilab, 2011.
\bibitem{bylinkin} A.A. Bylinkin(Moscow, MIPT) and O.I.
Piskounova (LPI), e-Print: arXiv:1501.07706.
\bibitem{lhcbmeson}LHCb Collaboration,JHEP {\bf 04}(2012)93, e-print - arXiv:1202.4812.
\bibitem{rhadron}Y.R. de Boer( Twente Uni.), A.B. Kaidalov( ITEP), D.A.Milstead (Stockholm Uni.) and O.I. Piskounova (LPI),J.Phys. G{\bf 35}(2008)075009, e-Print: arXiv 0710.3830.
\bibitem{charm}A.B.Kaidalov, O.I.Piskounova, Yad.Fiz. {\bf 43}(1986)1545, O.I.Piskounova,Phys.Atom.Nucl. {\bf 56} (1993)1094.
\bibitem{sigmap} O.I. Piskunova (Lebedev Inst.)Phys.Atom.Nucl. {\bf 64} (2001)98, O.I. Piskounova (Lebedev Inst.) Phys.Atom.Nucl. {\bf 66} (2003)307, e-Print: hep-ph/0202005.  
\bibitem{nikitin}O.I. Piskounova (Lebedev Inst.) and N.V. Nikitin (SINP, Moscow), Phys.Atom.Nucl. {\bf 68}(2005)2124, e-Print: hep-ph/0503006.  
\bibitem{extrabaryons} V.Greco, C.Ko, P.Levai Phys.Rev.Lett. {\bf 90}(2003)202302.
\bibitem{lhcb} LHCb Collaboration, Chin.Phys.C,C{\bf 40}(2016)011001, e-Print -  arXiv:1509.00292.
\bibitem{antiprotonratio}O.I.Piskunova (Lebedev Inst.), Sov.J.Nucl.Phys. {\bf 47}(1988)480.
\bibitem{beauty}O.I.Piskounova, Phys.Atom.Nucl.{\bf 64}(2001) 392, e-Print: hep-ph/0001252.
\bibitem{recent} Olga Piskounova, in the Proceedings of "38th International Conference of High Energy" PoS(ICHEP2016)711, e-print - arXiv:1602.08003.

\end{thebibliography}
\end{document}